\documentstyle[12pt,epsfig,amssymb]{article}
\newlength{\dinwidth}
\newlength{\dinmargin}
\newcommand {\beq}{\begin{equation}}
\newcommand {\eeq}{\end{equation}}
\newcommand {\bea}{\begin{eqnarray}}
\newcommand {\eea}{\end{eqnarray}}
\newcommand {\nn}{\nonumber}
\newcommand {\tr}{{\rm tr\,}}
\newcommand {\Tr}{\mbox{Tr\,}}

\newcommand {\dd}{\mbox{d}}

\newcommand {\del}{\partial}

\newcommand {\limit}{\rightarrow}

\newbox\pippobox

\begin{document}
\thispagestyle{empty} \addtocounter{page}{-1}
\begin{flushright}
OIQP-05-19\\
%
{\tt hep-lat/0601024}\\
\end{flushright} 
\vspace*{1cm}

\centerline{\Large \bf  Two-Dimensional Compact ${\cal N}=(2, 2)$}
\vskip0.4cm
\centerline{\Large \bf Lattice Super Yang-Mills Theory} 
\vskip0.4cm
\centerline{\Large \bf  with Exact Supersymmetry} 
\vspace*{2cm}
\centerline{\large Fumihiko Sugino} 
\vspace*{1.0cm}
\centerline{\it Okayama Institute for Quantum Physics} \vspace*{0.2cm}
\centerline{\it Kyoyama 1-9-1, Okayama 700-0015, Japan}
\vspace*{0.8cm}
\centerline{\tt fumihiko\_sugino@pref.okayama.jp} 
\vskip2cm
\centerline{\bf Abstract}
\vspace*{0.3cm}
{\small 

We construct two-dimensional ${\cal N}=(2,2)$ lattice super Yang-Mills theory, 
where the gauge and Higgs fields are all represented by U$(N)$ compact variables, 
with keeping one exact supercharge along the line of the papers \cite{sugino, sugino2, sugino3}. 
Interestingly, requirements of the exact supersymmetry as well as of the compact gauge and Higgs 
fields lead to the gauge group U$(N)$ rather than SU$(N)$. 
As a result of the perturbative renormalization argument, the model is shown to flow to 
the target continuum theory without any fine-tuning.   
Different from the case of noncompact Higgs fields, the path integral along 
the flat directions is well-defined in this model. 
}
\vspace*{1.1cm}



\newpage


\section{Introduction}

Lattice formulations of supersymmetric theories have a long history and still have been 
vigorously investigated \cite{nicolai}--\cite{elliott}\footnote{
For recent reviews, see refs. \cite{feo}.}. 
Recently, a formulation of super Yang-Mills (SYM) theories 
based on the idea of the deconstruction has been discussed \cite{kaplan}\footnote{
For related works, see refs. \cite{kaplan_related, kaplan_related2}.}. Also, various supersymmetric 
lattice models have been constructed from the connection to topological field theory 
via twisted supersymmetry \cite{sugino, sugino2, sugino3, catterall, catterall2, kawamoto}. 

In this paper, starting from the two-dimensional ${\cal N}=(2,2)$ SYM model 
constructed in ref.~\cite{sugino2}\footnote{
In the works \cite{sugino, sugino2, sugino3, sugino-conf}, 
we are considering nonchiral theories and using the term 
``two-dimensional ${\cal N}=2$ or 4", although it may be most precise to use 
``two-dimensional ${\cal N}=(2,2)$ or $(4,4)$" .}, 
we will present a new lattice model where the Higgs scalar fields are represented by compact variables 
with one supercharge exactly realized. In the previous model discussed in \cite{sugino2}, the 
Higgs fields are described as noncompact variables so that the flat directions (the minima of the Higgs potential) 
are continuously distributed over a noncompact region. It may cause some difficulty for the path integral over 
the configurations corresponding to the flat directions on performing actual 
numerical simulations. The model presented here is supersymmetric and free from the problem. 
In a sense, this model may be regarded as a modification of the model considered in ref.~\cite{suzuki-taniguchi} 
so that the supersymmetry $Q$ is exactly realized on the lattice. 
{}From the theoretical point of view, in theories defined on three- or higher dimensions, 
each of the Higgs vacuum 
expectation values in the flat directions composes a superselection sector.  
Because the superselection sectors do not affect each other, 
we can say that each of the Higgs vacuum expectation values defines a theory independently. 
On the other hand, in two-dimensional case, because continuous symmetries are not spontaneously 
broken \cite{mermin}, 
we should take into account all the superselection sectors together. 
Namely, we have to sum over all the contributions from the flat directions. 
In the lattice model constructed here, the summation over the flat directions is unambiguously defined, 
and its dynamics can be explicitly investigated with some supersymmetry preserved.   
We hope it useful for a progress on lattice formulations of 
supersymmetric theories, although the positivity of the fermion determinant 
remains to be investigated for numerical computations of the model. 
 
We consider the gauge group $G={\rm U}(N)$ instead of SU$(N)$. 
The case $G={\rm SU}(N)$ seems not 
to realize the exact supersymmetry with the compact gauge and Higgs fields in a straightforward manner. 
For $G={\rm U}(N)$, we impose some admissibility condition for the lattice action 
in order to resolve the problem of the vacuum degeneracy in the gauge terms 
as discussed in ref.~\cite{sugino2}\footnote{
The similar problem was encountered in ref. \cite{elitzur}.}. 
The action has a similar form to that of the U(1) chiral gauge theory presented by 
L\"uscher \cite{luscher}. Unfortunately, it does not allow to perform the strong coupling 
expansion with respect to the gauge part. 
It is quite interesting to overcome this point and modify the model to make  
the strong coupling expansion possible. 
    
This paper is organized as follows. In section~\ref{sec:2DN=2_noncompact}, we discuss the two-dimensional 
${\cal N}=(2, 2)$ SYM lattice model for the gauge group $G={\rm U}(N)$ introduced in \cite{sugino2}. 
Also, the path-integral measure is shown to be invariant under the exact supersymmetry $Q$. 
The renormalization argument for the $G={\rm U}(N)$ case is presented. 
In section~\ref{sec:2DN=2_compact}, we modify the supersymmetry $Q$ in the last section so that 
the Higgs fields can be represented as compact unitary variables. We construct the lattice action where 
all the gauge and Higgs fields are compact variables with the modified exact supersymmetry $Q$ preserved. 
The invariance of the path-integral measure is also maintained. 
The perturbative analysis shows that the target continuum theory is obtained from the lattice model 
without any fine-tuning. 
In this action, the flat directions are compactified to a finite domain, 
and thus the difficulty on the numerical simulations is resolved. 
We summarize the results obtained so far and discuss some future directions in section~\ref{sec:summary}. 
   

\section{Two-dimensional Lattice ${\cal N}= (2,2)$ SYM with Noncompact Higgs Fields} 
\label{sec:2DN=2_noncompact}
\setcounter{equation}{0}

We discuss the supersymmetric lattice action of two-dimensional ${\cal N}=(2,2)$ 
SYM theory for $G = {\rm U}(N)$ presented in \cite{sugino2}\footnote{It is 
based on a formulation of topological field theory in continuum space-time 
by Witten \cite{witten}.}. 

Other than the gauge link variables $U_{\mu}(x)\in {\rm U}(N)$ ($\mu=1, 2$), 
the two-dimensional ${\cal N}=(2,2)$ SYM theory has noncompact complex Higgs scalars 
$\phi(x)=X_3(x) +iX_4(x)$, $\bar{\phi}(x)=X_3(x)-iX_4(x)$, 
and fermions denoted as $\psi_{\mu}(x)$, $\chi(x)$, $\eta(x)$. 
They are transformed under the exact supersymmetry $Q$ as 
\bea
 & & QU_{\mu}(x) = i\psi_{\mu}(x) U_{\mu}(x),  \nn \\
 & & Q\psi_{\mu}(x) = i\psi_{\mu}(x)\psi_{\mu}(x) 
    -i\left(\phi(x) - U_{\mu}(x)\phi(x+\hat{\mu})U_{\mu}(x)^{-1}\right),
  \nn \\
 & & Q\phi(x) = 0,     \nn \\
 & & Q\chi(x) = H(x), \quad 
           QH(x) = [\phi(x), \,\chi(x)], \nn \\
 & & Q\bar{\phi}(x) = \eta(x), \quad  Q\eta(x) = [\phi(x), \,\bar{\phi}(x)],  
\label{Q_noncompact}
\eea
where $H(x)$ is an auxiliary field. Bosonic fields 
$H(x)$, $X_3(x)$, $X_4(x)$ and the fermions are $N\times N$ hermitian matrices put 
on the lattice site $x$. 
$Q$ is nilpotent up to the 
infinitesimal gauge transformation with the complexified parameter 
$\phi(x)$. In terms of $X_3(x)$ and $X_4(x)$, the transformation reads 
\beq
QX_3(x) = \frac12\eta(x), \quad QX_4(x) = \frac{i}{2}\eta(x). 
\label{QX}
\eeq 

Notice that the $Q$-transformation of $U_{\mu}(x)$ is a left-action of U$(N)$ group 
and remains the U$(N)$ Haar measure $[\dd U_{\mu}(x)]$ invariant. 
For the real Grassmann parameter $\varepsilon$ ($\varepsilon^2=0$), 
\beq
U_{\mu}(x) \mapsto e^{-\varepsilon \psi_{\mu}(x)}U_{\mu}(x) = U_{\mu}(x) + i\varepsilon QU_{\mu}(x). 
\eeq
Since $\varepsilon \psi_{\mu}(x)$ is anti-hermitian\footnote{
$(\varepsilon \psi_{\mu}(x))^{\dagger} = \psi_{\mu}(x)^{\dagger}\varepsilon^{\dagger}
=\psi_{\mu}(x)\varepsilon = -\varepsilon \psi_{\mu}(x).$}, 
$e^{-\varepsilon \psi_{\mu}(x)}$ may be regarded as an element of U$(N)$. 

In the expansion by a basis of the hermitian matrices $\{T^a\}_{a=1, \cdots, N^2}$: 
$({\rm field})(x)=\sum_a ({\rm field})^a(x) T^a$, 
the coefficients $\phi^a(x)$, $\bar{\phi}^a(x)$ are complex, and 
the fermionic variables $\psi_{\mu}^a(x)$, $\chi^a(x)$, $\eta^a(x)$ 
may be regarded as complexified Grassmann\footnote{Note that since the action is 
a holomorphic function with respect to the complexified Grassmann variables and 
there do not appear the anti-holomorphic variables, 
the degrees of freedom of the fermions do not change by the complexification. 
In a sense, it can be regarded as a sort of analytic continuation of 
the Grassmann variables~\cite{waldron}. } 
to be compatible to 
the U$(1)_R$ rotations (\ref{U1R}) given later. 
Notice that $\phi^a(x)$ and $\bar{\phi}^a(x)$ can 
be treated as independent variables in the path integral 
and that each of $H^a(x)$ is allowed to be 
shifted by a complex number. Thus, (\ref{Q_noncompact}) is seen to be consistently 
closed in the path-integral expression of the theory.  

The lattice action is constructed as  
\bea
S^{{\rm noncompact}}_{2D{\cal N}=2} & = & 
 Q\frac{1}{2g_0^2}\sum_x \, \tr\left[ 
\frac14 \eta(x)\, [\phi(x), \,\bar{\phi}(x)] -i\chi(x)\hat{\Phi}(x) 
+\chi(x)H(x)\right. \nn \\
 & & \hspace{2cm}\left. \frac{}{} 
+i\sum_{\mu=1}^2\psi_{\mu}(x)\left(\bar{\phi}(x) - 
U_{\mu}(x)\bar{\phi}(x+\hat{\mu})U_{\mu}(x)^{-1}\right)\right]
\label{S_2DN2}
\eea
when $||1-U_{12}(x)|| < \epsilon$ for $\forall x$, and 
\beq
S^{{\rm noncompact}}_{2D{\cal N}=2} = + \infty 
\label{S_2DN2_2}
\eeq
otherwise.  
Here $U_{\mu\nu}(x)$ are plaquette variables written as
\beq
U_{\mu\nu}(x) \equiv U_{\mu}(x) U_{\nu}(x+\hat{\mu}) 
U_{\mu}(x+\hat{\nu})^{-1} U_{\nu}(x)^{-1} = U_{\nu\mu}(x)^{\dagger}. 
\eeq
For definiteness, here and in what follows we use the following 
definition of the norm of a matrix $A$:   
\beq
||A|| \equiv \left[\tr \left(AA^{\dagger}\right)\right]^{1/2}, 
\label{norm}
\eeq
and $\epsilon$ is a positive number chosen as 
\beq
0 < \epsilon < 2.  
\label{epsilon} 
\eeq
Also 
\bea
\hat{\Phi}(x) & =  & \frac{\Phi(x)}{1-\frac{1}{\epsilon^2}||1-U_{12}(x)||^2},  
\label{Phihat_2d} \\
\Phi(x) & =  & -i\left[U_{12}(x)- U_{21}(x)\right].    
\label{Phi_2d} 
\eea

Explicitly, (\ref{S_2DN2}) is 
\bea
S^{{\rm noncompact}}_{2D{\cal N}=2} & = & \frac{1}{2g_0^2}\sum_x \, \tr\left[
\frac14 [\phi(x), \,\bar{\phi}(x)]^2 + H(x)^2 
-iH(x)\hat{\Phi}(x) \right. \nn \\
 & & 
+\sum_{\mu=1}^2\left(\phi(x)-U_{\mu}(x)\phi(x+\hat{\mu})U_{\mu}(x)^{\dagger}
\right)\left(\bar{\phi}(x)-U_{\mu}(x)\bar{\phi}(x+\hat{\mu})
U_{\mu}(x)^{\dagger}\right) \nn \\
 & & -\frac14 \eta(x)[\phi(x), \,\eta(x)] 
- \chi(x)[\phi(x), \,\chi(x)] \nn \\
 & & 
-\sum_{\mu=1}^2\psi_{\mu}(x)\psi_{\mu}(x)\left(\bar{\phi}(x)  + 
U_{\mu}(x)\bar{\phi}(x+\hat{\mu})U_{\mu}(x)^{\dagger}\right) \nn \\
 & & \left. \frac{}{}+ i\chi(x) Q\hat{\Phi}(x) 
-i\sum_{\mu=1}^2\psi_{\mu}(x)\left(\eta(x)-
U_{\mu}(x)\eta(x+\hat{\mu})U_{\mu}(x)^{\dagger}\right)\right]. 
\label{S_2DN2_explicit}
\eea
After integrating out $H(x)$, the gauge kinetic terms are induced as 
\beq
\frac{1}{8g_0^2}\sum_x\,\tr \left(\hat{\Phi}(x)^2\right) 
= \frac{1}{8g_0^2}\sum_x\,
   \frac{\tr \left[2-U_{12}(x)^2-U_{21}(x)^2\right]}{
    \left(1-\frac{1}{\epsilon^2}||1-U_{12}(x)||^2\right)^2}. 
\eeq
Although the numerator means the minima given by the configurations 
satisfying $U_{12}(x)^2=1$, the admissibility condition 
$||1-U_{12}(x)|| < \epsilon$ with (\ref{epsilon}) allows the unique one $U_{12}(x)=1$ 
among them. The denominator guarantees smoothness 
of the Boltzmann weight $\exp\left[-S^{{\rm noncompact}}_{2D{\cal N}=2}\right]$. 
Also, the Boltzmann weight consists of a product of local factors, leading the locality of 
the theory.     
The form of the action is somewhat 
similar to that of the U(1) chiral gauge theory constructed by 
L\"{u}scher \cite{luscher}. 

The action (\ref{S_2DN2}) is clearly $Q$-invariant 
from its $Q$-exact form. 
Furthermore, the invariance under the following global U$(1)_R$ rotation 
holds: 
\bea
U_{\mu}(x) \limit U_{\mu}(x), & & 
\psi_{\mu}(x) \limit e^{i\alpha}\psi_{\mu}(x), \nn \\ 
\phi(x) \limit e^{2i\alpha}\phi(x), &  &  \nn \\
\chi(x) \limit e^{-i\alpha}\chi(x), & & H(x) \limit H(x), \nn \\
\bar{\phi}(x) \limit e^{-2i\alpha}\bar{\phi}(x), & & 
\eta(x) \limit e^{-i\alpha}\eta(x). 
\label{U1R}
\eea
The U$(1)_R$ charge of each variable is read off from (\ref{U1R}), and 
$Q$ increases the charge by one unit. 

We can see the $Q$-invariance of the path-integral measure of the model:
\beq
\dd \mu^{{\rm noncompact}}_{2D{\cal N}=2} \equiv \left(\prod_{\mu}[\dd U_{\mu}(x)] \right)
[\dd X_3(x)] [\dd X_4(x)] [\dd H(x)] 
     \left(\prod_{\mu} [\dd \psi_{\mu}(x)] \right)[\dd \chi(x)] [\dd \eta(x)]
\eeq
with  
$[\dd ({\rm field})(x)] \equiv \prod_x\prod_{a=1}^{N^2}\dd ({\rm field})^a(x)$ 
for each hermitian field. 
If we express the hermitian fields collectively as $\varphi_I$ with the index $I$ 
running over the species of the fields, the sites and the gauge indices: 
\beq
\varphi_I =\{ X_3^a(x), X_4^a(x), H^a(x), \psi_{\mu}^a(x), \chi^a(x), \eta^a(x) \}, 
\eeq       
the response of the measure under the $Q$-transformation is written as 
\beq
\dd\mu^{{\rm noncompact}}_{2D{\cal N}=2} \limit \dd\mu^{{\rm noncompact}}_{2D{\cal N}=2} 
    \left(1+ i\varepsilon \sum_I (-1)^{|\varphi_I|} \, \partial (Q\varphi_I)/\partial \varphi_I\right).  
\label{jacobian}
\eeq
$(-1)^{|\varphi_I|}$ means the statistics of the field $\varphi_I$: $+1$ ($-1$) for bosons (fermions). 
Although $QU_{\mu}(x)$ contains the variable $\psi_{\mu}(x)$, the effect to the measure 
is proportional to $\varepsilon^2(=0)$ because it contributes to off-diagonal parts of the 
Jacobian matrix.  
Note that $\partial (Q\psi_{\mu}^a(x))/\partial\psi_{\mu}^a(x) =0$, 
since the first term of $Q\psi_{\mu}(x)$ in (\ref{Q_noncompact}) is 
\beq
i\psi_{\mu}(x)\psi_{\mu}(x) = -\frac{1}{2}\sum_{a,b,c}f^{abc}\psi_{\mu}^a(x)\psi_{\mu}^b(x)T^c 
\eeq
with $f^{abc}$ being the structure constant of U$(N)$. 
Thus, it is easily seen that 
$$\sum_I (-1)^{|\varphi_I|} \, \partial (Q\varphi_I)/\partial \varphi_I=0$$ 
and the measure $\dd\mu^{{\rm noncompact}}_{2D{\cal N}=2}$ is $Q$-invariant. 


\paragraph{Renormalization}
It is straightforward to see that after the rescaling as
\bea
 & & \phi(x) \limit a\phi(x), \quad \bar{\phi}(x) \limit a \bar{\phi}(x), \quad 
H(x) \limit a^2H(x), \quad  
\psi_{\mu}(x) \limit a^{3/2}\psi_{\mu}(x), \nn \\
 & & \chi(x) \limit a^{3/2}\chi(x), \quad \eta(x) \limit a^{3/2}\eta(x)  
\eea 
for the lattice spacing $a$, 
the lattice action (\ref{S_2DN2_explicit}) reduces to the continuum action of ${\cal N}=(2, 2)$ 
SYM in the continuum limit $a\limit 0$ with $g_2\equiv g_0/a$ fixed. 
Thus the full ${\cal N}=(2,2)$ supersymmetry and the rotational symmetry 
in two dimensions are restored in the classical sense. 
We will check whether the symmetry restoration persists against 
the quantum corrections, i.e. whether symmetries of the lattice action 
forbid any induced relevant or marginal operators which are to obstruct the symmetry 
restoration.  
In order to consider the quantum effects near the continuum limit, 
we assume the fixed point at $g_0=0$, which is suggested by the asymptotic freedom of the theory, 
and treat the quantum fluctuations in the perturbative way around $g_0=0$.  

It is useful for the renormalization argument 
to note symmetries of the lattice action (\ref{S_2DN2}): 
\begin{itemize}
\item lattice translational symmetry 
\item U($N$) gauge symmetry 
\item supersymmetry $Q$ 
\item global U$(1)_R$ internal symmetry\footnote{It is not anomalous because all the matter 
fields belong to the adjoint representation. 
Note that the first Chern class $\Tr_{{(\rm Adj)}} F_{12}^{{(\rm Adj)}}$ with respect to the adjoint 
representation vanishes. } 
\item reflection symmetry  
$x\equiv (x_1, x_2) \limit \tilde{x}\equiv (x_2, x_1)$ with 
\bea
(U_1(x), U_2(x)) & \limit & (U_2(\tilde{x}), U_1(\tilde{x})) \nn \\
(\psi_1(x), \psi_2(x)) & \limit & (\psi_2(\tilde{x}), \psi_1(\tilde{x})) \nn \\
(H(x), \chi(x)) & \limit & (-H(\tilde{x}), -\chi(\tilde{x})) \nn \\
(\phi(x), \bar{\phi}(x), \eta(x) ) & \limit &  (\phi(\tilde{x}), \bar{\phi}(\tilde{x}), \eta(\tilde{x})). 
\label{reflection}
\eea
\end{itemize}

The mass dimension of the coupling constant squared $g_2^2$ is two. 
For generic boson field $\varphi$ (other than the auxiliary field) 
and fermion field $\psi$, the dimensions 
are 1 and 3/2 respectively.   
Thus, operators of the type $\varphi^a \del^b\psi^{2c}$ 
have the dimension $p\equiv a+b+3c$, where 
`$\del$' means a derivative with respect to a coordinate. 
{}From the dimensional analysis, the operators receive the following 
radiative corrections up to some powers of possible logarithmic factors:    
\beq
\left(\frac{a^{p-4}}{g_2^2} + c_1 a^{p-2} + c_2 a^{p}g_2^2 + \cdots\right)
\int \dd^2x\,  \varphi^a \del^b \psi^{2c},  
\label{loop_correction_2d}
\eeq 
where the first term in the parentheses represents the contribution from the tree level, 
and the term containing the coefficient $c_{\ell}$ comes from the $\ell$-loop contributions. 
It is easily seen from $g_2^2$ playing the same role as the Planck constant $\hbar$ 
in the action (\ref{S_2DN2}). From the formula (\ref{loop_correction_2d}), 
we read that operators with $p=1, 2$ are relevant or marginal 
arising only at the one- and two- loop levels. 

Because we know that the lattice action reduces to the desired continuum SYM action in the classical 
continuum limit, we should check operators with $p=0, 1, 2$. Operators with $p\leq 4$ are listed 
in Table~\ref{tab:operators}. The identity operator corresponding to $p=0$ 
merely shifts the action by a constant, 
which is not interesting to us.   
%
\begin{table}
\begin{center}
\begin{tabular}{|c|ccccc|}
\hline \hline
$p=a+b+3c$ & 
\multicolumn{5}{|c|}{$\varphi^a\del^b\psi^{2c}$}  \\ 
\hline
0 &     &          &  1       &       &    \\
1 &     &          & $\varphi$  &      &     \\
2 &     &          & $\varphi^2$ &     &     \\
3 &     &$\varphi^3$, & $\psi\psi$, & $\varphi\del\varphi$  &     \\
4 & $\varphi^4$, & $\varphi^2\del\varphi$, & $(\del\varphi)^2$, & 
   $\psi\del\psi$, & $\varphi\psi\psi$ \\
\hline \hline
\end{tabular}
\end{center}
  \caption{List of operators with $p\leq 4$. 
}
\label{tab:operators}
\end{table}
%
For the cases $p=1, 2$, the U($N$) gauge invariance and the U$(1)_R$ symmetry 
allow the scalar mass operator $\tr (\phi\bar{\phi})$ and the 
auxiliary field $\tr H$.  The former is 
forbidden by the supersymmetry $Q$, and the latter by the 
reflection symmetry (\ref{reflection}). 
Hence, it is seen that no relevant or marginal operators except the identity  
are generated by the radiative corrections, and that the model flows to the desirable 
continuum theory without any fine-tuning. 
As a conclusion, the rotational symmetry and the full ${\cal N}=(2, 2)$ supersymmetries are 
restored in the continuum limit.  

Since the Higgs fields $\phi(x)$ and $\bar{\phi}(x)$ are noncompact, the action has 
noncompact flat directions satisfying $[\phi(x), \bar{\phi}(x)]=0$. 
Numerical calculations for the model may lead to divergent quantities from the integration along 
the flat directions.  
In the next section, 
we consider a modification to the model where the Higgs fields are compactified 
with keeping the supersymmetry $Q$. 
The new model is supersymmetric and resolves the problem of the flat directions.

\section{Two-dimensional Lattice ${\cal N}= (2,2)$ SYM with Compact Higgs Fields} 
\label{sec:2DN=2_compact}
\setcounter{equation}{0}

In this section, we promote the variables $X_3(x)$ and $X_4(x)$ 
to the U$(N)$ matrices: 
\beq
X_3(x) \limit V_3(x)\equiv e^{iaX_3(x)}, \qquad 
X_4(x) \limit V_4(x)\equiv e^{iaX_4(x)}   
\eeq
sitting on the lattice site $x$, 
and construct a supersymmetric lattice theory where all the bosonic variables 
except the auxiliary field $H(x)$ are represented as compact unitary matrices. 

As counterparts of noncompact $\phi(x)$ and $\bar{\phi}(x)$, let us introduce   
\bea
V(x) & \equiv & -i(V_3(x) +iV_4(x)) = 1-i +a \phi(x) + O(a^2), \nn \\
V(x)^{\dagger} & = & i(V_3(x)^{-1} -iV_4(x)^{-1}) = 1+i +a \bar{\phi}(x) + O(a^2). 
\eea
As indicated in the r.h.s., they reproduce $\phi(x)$, $\bar{\phi}(x)$ for small $a$ 
up to some unimportant additive constants. 
Similarly to the transformation $QU_{\mu}(x)$ in the last section, 
we would like to consider a suitable modification of the $Q$-transformation 
which remains the Haar measures $[\dd V_3(x)]$ and $[\dd V_4(x)]$ invariant 
with keeping the nilpotency 
(up to infinitesimal gauge transformation). 
Suppose that such a transformation can be written as left- or right- actions of U$(N)$ group: 
\bea
V_3(x) & \mapsto & e^{i\varepsilon A(x)}V_3(x) e^{i\varepsilon B(x)} 
     = V_3 + i\varepsilon (A(x)V_3(x)+V_3(x)B(x)), \nn \\
V_4(x) & \mapsto & e^{i\varepsilon E(x)}V_4(x) e^{i\varepsilon F(x)} 
     = V_4(x) + i\varepsilon (E(x)V_4(x)+V_4(x)F(x)), 
\label{ABEF}     
\eea
with a real Grassmann parameter $\varepsilon$. 
Also, $A(x)$ and $B(x)$ ($E(x)$ and $F(x)$) should be Grassmann odd and independent 
of $V_3(x)$ ($V_4(x)$). 
The $Q$-transformation reads as
\beq
QV_3(x) = A(x)V_3(x) + V_3(x)B(x), \qquad QV_4(x) = E(x)V_4(x) + V_4(x) F(x) 
\label{QV}
\eeq
Here, we demand $QV(x)=0$ corresponding to $Q\phi(x)=0$ in (\ref{Q_noncompact}), 
and require that (\ref{QV}) reduces to (\ref{QX}) in the $a\limit 0$ limit.
As a solution, we shall take a choice of $A(x)$, $B(x)$, $E(x)$, $F(x)$ as\footnote{
Of course, there are other solutions for $A(x)$, $B(x)$, $E(x)$, $F(x)$ satisfying the requirements. 
For example, we can choose 
\beq
A(x) = 0, \quad B(x) = \frac{i}{2}\eta(x) V_4(x), \quad E(x) = -\frac12 V_3(x)\eta(x), \quad F(x) = 0   
\label{ABEF_sol2}
\eeq
instead of (\ref{ABEF_sol}). The symmetrized version 
\beq
A(x) = \frac{i}{4}V_4(x)\eta(x), \quad B(x) = \frac{i}{4}\eta(x) V_4(x), 
   \quad E(x) = -\frac14 V_3(x)\eta(x), \quad F(x) = -\frac14\eta(x)V_3(x) 
\label{ABEF_sol3}
\eeq
is also a solution. 
In a later analysis, however it turns out that (\ref{ABEF_sol3}) does not lead to consistently 
closed $Q$-algebra.}   
\beq
A(x)= \frac{i}{2}V_4(x)\eta(x), \quad B(x)=E(x)=0, \quad F(x)=-\frac12\eta(x) V_3(x). 
\label{ABEF_sol}
\eeq
Then, 
\bea
QV_3(x)^{-1} & = & -V_3(x)^{-1}(QV_3(x))V_3(x)^{-1} = -\frac{i}{2}V_3(x)^{-1}V_4(x)\eta(x), \nn \\
QV_4(x)^{-1} & = & -V_4(x)^{-1}(QV_4(x))V_4(x)^{-1} = \frac12\eta(x) V_3(x)V_4(x)^{-1}, \nn \\
QV(x)^{\dagger} & = & 
               \frac12 V_3(x)^{-1}V_4(x)\eta(x) + \frac12\eta(x) V_3(x)V_4(x)^{-1}. 
\eea
  
Note that, due to the Grassmann nature of $\varepsilon$ in (\ref{ABEF}), 
$A(x)$, $B(x)$, $E(x)$, $F(x)$ cause 
only infinitesimal actions to the U$(N)$ group manifold. 
Although $\varepsilon A(x)$ and $\varepsilon F(x)$ are not hermitian in (\ref{ABEF_sol}), they  
are expanded by a basis of $N\times N$ hermitian matrices with complex coefficients. 
Thus, the actions $\varepsilon A(x)$ and $\varepsilon F(x)$  
can be regarded as some infinitesimal complexified translations, 
which is somewhat similar to the infinitesimal 
complexified gauge transformation with the parameter $\phi(x)$ induced from the $Q$-supersymmetry 
in the last section.  

Here let us comment on the $G={\rm SU}(N)$ case. 
If we consider the case $G={\rm SU}(N)$ instead of U$(N)$, 
we must further impose the traceless condition on $A(x)$, $B(x)$, $E(x)$, $F(x)$. 
It seems quite nontrivial how to do it. In this paper, we do not consider it and 
concentrate to the U$(N)$ case. 

Next, in order to close the $Q$-algebra, we impose 
\beq
Q^2 V(x)^{\dagger}=[V(x), V(x)^{\dagger}], 
\label{Q2V}
\eeq
which determines $Q\eta(x)$ as 
\beq
Q\eta(x) = \frac{i}{2}\eta(x) V(x)\eta(x) +2i\left(V_4(x)^{-1}V_3(x)V_4(x)V_3(x)^{-1} -1\right). 
\label{Qeta}
\eeq
Then, happily we can check the closure as\footnote{
For the solution (\ref{ABEF_sol2}) we can also find the transformation rule $Q\eta(x)$ 
consistent with (\ref{Q2V}) as 
\bea
Q\eta(x) & = & -\frac{i}{2}\eta(x) V(x)\eta(x) 
            -2i\left(V_3(x)^{-1}V_4(x)V_3(x)V_4(x)^{-1} - 1\right),  \nn \\
Q^2\eta(x) & = & [V(x), \eta(x)], 
\eea
while for (\ref{ABEF_sol3}) we can not.}
$Q^2\eta(x) =[V(x), \eta(x)]$. 

Finally, the $Q$-transformation of the compact model is summarized as 
\bea
 & & QU_{\mu}(x) = i\psi_{\mu}(x) U_{\mu}(x),  \nn \\
 & & Q\psi_{\mu}(x) = i\psi_{\mu}(x)\psi_{\mu}(x)
         -i\left(V(x)-U_{\mu}(x)V(x+\hat{\mu})U_{\mu}(x)^{-1}\right), \nn \\
 & & QV(x)=0, \nn \\
 & & Q\chi(x)=H(x), \quad QH(x) = [V(x), \chi(x)], \nn \\
 & & QV(x)^{\dagger} = \frac12V_3(x)^{-1}V_4(x)\eta(x) + \frac12\eta(x) V_3(x)V_4(x)^{-1}, \nn \\
 & & Q\eta(x) = \frac{i}{2}\eta(x) V(x)\eta(x) + 2i\left(V_4(x)^{-1}V_3(x)V_4(x)V_3(x)^{-1}-1\right), 
\label{Q_compact} 
\eea
where 
\beq
QV_3(x) = \frac{i}{2}V_4(x)\eta(x) V_3(x),   \quad QV_4(x) = -\frac12 V_4(x)\eta(x) V_3(x). 
\eeq
It is nilpotent up to the infinitesimal gauge transformation with the complexified parameter 
$V(x)$. After expanding $V_3(x)$ and $V_4(x)$ for small $a$, 
(\ref{Q_compact}) reduces to the noncompact case 
(\ref{Q_noncompact}) (with the rescaling $\phi(x)\limit a^{-1}\phi(x)$, 
$\bar{\phi}(x) \limit a^{-1}\bar{\phi}(x)$). 
 
Making use of the transformation rule (\ref{Q_compact}), 
we write down the action of the $Q$-exact form as 
\bea
S^{{\rm compact}}_{2D{\cal N}=2} & = & Q\frac{1}{2g_0^2}\sum_x \,\tr 
        \left[-\frac{i}{2}\eta(x) \left(V_3(x)V_4(x)^{-1}V_3(x)^{-1}V_4(x)-1\right) 
             -i\chi(x)\hat{\Phi}(x) \right. \nn \\
  &   &  
\left. \frac{}{} +\chi(x)H(x) 
         + i\sum_{\mu =1}^2\psi_{\mu}(x)
              \left(V(x)^{\dagger} - U_{\mu}(x)V(x+\hat{\mu})^{\dagger}U_{\mu}(x)^{-1}\right)
                            \right]
\label{S_2DN2_compact}
\eea
when $||1-U_{12}(x)|| < \epsilon$ for $\forall x$, and 
\beq
S^{{\rm compact}}_{2D{\cal N}=2} = + \infty 
\label{S_2DN2_compact_2}
\eeq
otherwise, 
with $\hat{\Phi}(x)$ chosen as (\ref{Phihat_2d}).   
After the $Q$-action in (\ref{S_2DN2_compact}), the action becomes 
\bea
S^{{\rm compact}}_{2D{\cal N}=2} & = & 
            \frac{1}{2g_0^2} \sum_x \,\tr \left[-\frac14 \eta(x) V(x) \eta(x) 
                         \left(V_3(x)V_4(x)^{\dagger}V_3(x)^{\dagger}V_4(x)+1\right) \right. \nn \\
  &   &     + [V_3(x), V_4(x)][V_4(x)^{\dagger}, V_3(x)^{\dagger}] +H(x)^2 -iH(x)\hat{\Phi}(x) \nn \\
  &   &    +i\chi(x)\left(Q\hat{\Phi}(x)\right) -\chi(x) [V(x), \chi(x)] \nn \\
  &   &   -\sum_{\mu = 1}^2 \psi_{\mu}(x)\psi_{\mu}(x)
                      \left(V(x)^{\dagger} + U_{\mu}(x)V(x+\hat{\mu})^{\dagger}U_{\mu}(x)^{\dagger}\right) \nn \\
  &   &  +\sum_{\mu = 1}^2 \left(V(x) - U_{\mu}(x)V(x+\hat{\mu})U_{\mu}(x)^{\dagger}\right)
                           \left(V(x)^\dagger - U_{\mu}(x)V(x+\hat{\mu})^{\dagger}U_{\mu}(x)^{\dagger}\right) \nn \\
  &   &  -\frac{i}{2}\sum_{\mu =1}^2 \psi_{\mu}(x)\left(
                              V_3(x)^{\dagger}V_4(x)\eta(x) +\eta(x) V_3(x)V_4(x)^{\dagger} \right. \nn \\
  &   & \hspace{2.5cm}   -U_{\mu}(x)V_3(x+\hat{\mu})^{\dagger}V_4(x+\hat{\mu})\eta(x+\hat{\mu}) U_{\mu}(x)^{\dagger} \nn \\
  &   & \hspace{2.3cm}     \left. \frac{}{} \left.
                       - U_{\mu}(x)\eta(x+\hat{\mu}) V_3(x+\hat{\mu})V_4(x+\hat{\mu})^{\dagger}U_{\mu}(x)^{\dagger}
                                          \right)\right]. 
\eea
In this action, the Higgs fields are expressed by the compact variables $V_3(x)$ and $V_4(x)$, and thus 
the flat directions satisfying $[V_3(x), V_4(x)]=0$ are compactified to a finite region 
with the supersymmetry $Q$ maintained.  
After $H(x)$ integrated out, $\hat{\Phi}^2$ term is induced: 
\beq
\frac{1}{8g_0^2}\sum_x\, \tr \left(\hat{\Phi}(x)^2\right) = \frac{1}{8g_0^2}\sum_x
 \frac{\tr(2-U_{12}(x)^2-U_{21}(x)^2)}{(1-\frac{1}{\epsilon^2}||1-U_{12}(x)||^2)^2}.
\label{Phisquare}
\eeq
We should remark that the strong coupling expansion is not possible with respect to (\ref{Phisquare}) due to 
the denominator\footnote{
Of course, it is impossible to perform the strong coupling expansion 
for the $Q$-exact action (\ref{S_2DN2_compact}) 
from the reasons both of the noncompact $H(x)$-integral and the denominator of $\hat{\Phi}(x)$. 
It gives a way out of Neuberger's no go theorem \cite{neuberger}.}. 
In $\exp(-S^{{\rm compact}}_{2D{\cal N}=2})$, 
the zeros of the denominator are essential singularities. It is essentially same 
as the following fact. The function 
\beq
f(t) = \left \{ \begin{array}{cc} \frac{1}{t^n}e^{-c/t^2} & 
\mbox{ for } t>0 \\ 0 & \mbox{ for } t \leq 0 \end{array} \right. 
\label{ft}
\eeq
with $c$ positive constant is smooth and differentiable at $t=0$ for 
$n=0, 1, 2, \cdots$. In evaluating the integral 
\beq
\int_{-L}^L \dd t \, f(t), 
\eeq
however it is not allowed to expand the exponential in (\ref{ft}) 
and integrate term-by-term. 

Interestingly, the U$(1)_R$ charge is still consistently defined in the compact model. 
The charge 0 is assigned to $U_{\mu}(x)$ and $H(x)$, $+1$ to $\psi_{\mu}(x)$, 
$-1$ to $\chi(x)$ and $\eta(x)$, 
$+2$ to $V_3(x)$ and $V_4(x)$, $-2$ to $V_3(x)^{-1}$ and $V_4(x)^{-1}$.  

We shall see the $Q$-invariance of the path-integral measure 
\beq
\dd \mu^{{\rm compact}}_{2D{\cal N}=2} \equiv \left(\prod_{\mu}[\dd U_{\mu}(x)] \right)
[\dd V_3(x)] [\dd V_4(x)] [\dd H(x)] 
     \left(\prod_{\mu} [\dd \psi_{\mu}(x)] \right)[\dd \chi(x)] [\dd \eta(x)]. 
\eeq
For the collective expression of the hermitian fields 
\beq
\varphi_I =\{ H^a(x), \psi_{\mu}^a(x), \chi^a(x), \eta^a(x) \}, 
\eeq  
the similar formula to (\ref{jacobian}) holds. Although $QV_3(x)$ contains $V_4(x)$, $\eta(x)$ 
and $QV_4(x)$ depends on $V_3(x)$ and $\eta(x)$, the effect to the measure vanishes again since 
$\varepsilon^2=0$. 
The potentially dangerous part is only the first term of $Q\eta(x)$ in (\ref{Q_compact}). 
Assuming the normalization of the basis $\tr (T^aT^b)=\delta^{ab}$ and the completeness 
$\sum_{a=1}^{N^2}(T^a)_{ij}(T^a)_{kl}=\delta_{il}\delta_{jk}$, 
\bea
\sum_a \partial(Q\eta^a(x))/\partial\eta^a(x) & = & 
               \frac{i}{2}\sum_a \,\tr\left(T^aT^a[\eta(x), V(x)]\right) \nn \\
    & = & \frac{i}{2}N\,\tr\left([\eta(x), V(x)]\right) = 0. 
\eea
Thus, $\sum_I (-1)^{|\varphi_I|} \, \partial (Q\varphi_I)/\partial \varphi_I=0$ leading the $Q$-invariance of 
the measure $\dd \mu^{{\rm compact}}_{2D{\cal N}=2}$.     

Clearly from the construction, the action (\ref{S_2DN2_compact}) reduces to the noncompact case 
(\ref{S_2DN2}) in the $a\limit 0$ limit. 
As long as considering the renormalization in the perturbative framework, 
the ultraviolet property of the model does not change from the noncompact case. 
Thus, the argument goes parallel to 
the previous section showing that the lattice model leads to the target continuum theory 
without any fine-tuning.  
  
\setcounter{equation}{0}
\section{Summary and Discussions}
\label{sec:summary}

In this paper, we have presented a lattice formulation of two-dimensional 
${\cal N}=(2,2)$ SYM theory, where 
the gauge and Higgs fields are all represented as U$(N)$ compact variables 
with keeping one exact supercharge $Q$. 
In this construction, the path-integral measure is shown to be $Q$-invariant. 
On the basis of the perturbative argument, it can be shown that the target continuum theory 
is obtained with no fine-tuning, similarly to the noncompact Higgs case. 
Here, we should remark that it does not exclude the possibility that 
the compact Higgs model may belong to a different 
universality class from the noncompact one due to some nonperturbative effects\footnote{
As such an example, the XY model 
on the two-dimensional lattice is well-known. 
The target space of the variable is a circle, and it is compared to a free scalar field taking values 
on an infinite real line.  
They exhibit the same behavior within the perturbative analysis of the XY model 
around the trivial vacuum. 
However, 
when taking into account the contribution from vortex configurations nonperturbatively, 
it turns out that the XY model takes place the Kosterlitz-Thouless phase transition 
at the nonzero coupling 
and has a different phase diagram from a free scalar field. 
For example, see ref.~\cite{kogut}.}.     
It will be important to perform nonperturbative investigation for both lattice models and 
clarify this point.  
 
In the case of noncompact Higgs fields discussed in the previous 
papers \cite{sugino, sugino2, sugino3}, the flat directions 
are continuously distributed and spread over noncompact domains. 
It may give divergent quantities from the path integrals along the flat directions 
on actual numerical simulations of the models. 
However, the model constructed in this paper has compact flat directions, and thus overcomes 
such obstruction with preserving the exact supersymmetry $Q$. 
 
Somewhat interestingly, 
in order to realize both of the compact gauge and Higgs fields and the exact supersymmetry, 
we have been led to 
consider the gauge group $G={\rm U}(N)$ rather than SU$(N)$.  
For the $G={\rm U}(N)$ case, we have employed a formulation based on the admissibility condition to 
resolve the problem of the degenerate vacua as discussed in \cite{sugino2}. As a result, 
the strong coupling expansion is not allowed in this formulation for the gauge terms. 
It is certainly worth to reformulate the theory so that the strong coupling expansion is 
possible with the exact supersymmetry preserved. If it succeeds, 
it could become an important step towards to investigate 
the strong coupling dynamics of SYM theories in the framework of supersymmetric lattice models.    

It is quite interesting to consider to extend the formulation to the other cases 
of two-dimensional ${\cal N}=(4,4), (8,8)$ and three-dimensional ${\cal N}=4, 8$ 
discussed in \cite{sugino, sugino2, sugino3}. 
In particular, for the cases of the two exact supercharges $Q_{\pm}$ realized, it is 
intriguing to find how to modify the lattice supersymmetry transformation rules so as to 
incorporate compact Higgs scalars.  
Since the models with 
two exact supercharges have much more rich symmetries compared to the cases of one supercharge, 
the realization is expected to be fruitful from both of theoretical and practical aspects of 
lattice field theories.




\end{document}